\def\noi{\noindent}
\def\pref#1{(\ref{#1})}
\def\beq{\begin{equation}}
\def\eeq{\end{equation}}
\def\bmulteq{\begin{eqnarray}}
\def\emulteq{\end{eqnarray}}

\def\CPV{$CP$ violation}
\def\CPvio{$CP$ violation}
\def\susy{supersymmetry}

\def\susyic{supersymmetric}
\def\Susyic{Supersymmetric}

\def\dn{$d_n$}

\def\ecm{e$\,$cm}

\def \Im{{\rm Im}}

\def \Arg{{\rm Arg}}

\def \vev#1{\left< #1 \right>}

\def\Lagr{Lagrangian}

\def\abs #1{|#1|}

\def\H{{\rm H}}

\def\plet #1 {\left\lgroup \matrix{#1} \right\rgroup}

\def\N{{\rm N}}

\documentstyle[preprint,revtex]{aps}

\begin{document}
\draft
\preprint{TRI-PP-1}
\preprint{Jan. 1993}
\begin{title}
What is the Natural Size of Supersymmetric $CP$ Violation?
\end{title}
\author{Robert Garisto}
\begin{instit}
\begin{center}
TRIUMF,
4004 Wesbrook Mall,
Vancouver, B.C.,
V6T 2A3,
Canada
\end{center}
\end{instit}
\author{Gordon Kane}
\begin{instit}
\begin{center}
Dept. of Physics, University of Michigan,
Ann Arbor, MI \ 48109, U.S.A.
\end{center}
\end{instit}
\begin{abstract}
It is well known that if phases and masses in the
Minimal Supersymmetric Standard Model (MSSM) are allowed to have general
values,
the resulting neutron EDM ($d_n$) exceeds the experimental upper limit by
about $10^3$.  We assume that the needed suppression is not due to
a fine-tuning of phases or masses, and ask what natural size of $CP$ violation
(CPV) results. We show that
(1) the phase of one of the superpotential parameters, $\mu$, does not
contribute to any CPV in the MSSM and so is not constrained by \dn;
(2) the MSSM contribution to $d_n$ is tiny, just coming from the CKM phase;
(3) the phases in the MSSM cannot be used to generate a baryon asymmetry
at the weak scale, given our assumptions; and
(4) in non-minimal SUSY models, an effective phase can enter at one loop
giving $d_n \sim 10^{-26}$\ecm, $d_e \sim 10^{-27}$\ecm, and allowing
a baryon asymmetry  to be generated at the weak scale, without fine-tunings.
Our results could be evaded by a SUSY breaking mechanism which
produced phases for the SUSY breaking parameters that somehow were naturally
of order $10^{-3}$.
\end{abstract}



\newpage
\section {Introduction}

Predictions for $CP$ violating effects in supersymmetric (SUSY) theories
have often been discussed with a certain ambiguity.
On the one hand, it is well known that when
the complex quantities in the theory are allowed to
have phases of order unity, the predicted neutron electric dipole
moment (\dn) is typically too large by perhaps $10^{3}$
\cite{hist dn,recent dn,arnowitt}.
In order to avoid this, the relevant quantities are often chosen
to be real, in which case the theory predicts no
non-Standard Model $CP$ violation or \dn\ at all.
On the other hand, it has often been assumed that
the observation of \dn\ around the current limit of $10^{-25}$\ecm\
\cite{dnexpt} could easily be accommodated by a SUSY theory with
the phases somehow reduced by just the right amount.
These ideas are clearly in conflict: one cannot have a theory which
avoids fine-tunings by having \dn\ zero, and at the same time which
gives \dn\ near the current limit.

All of the \CPV\ (CPV) induced by \susy\ occurs because of phases
in either the superpotential
or the soft SUSY breaking Lagrangian
\cite{foot strong}.
The superpotential of the Minimal \Susyic\ Standard Model (MSSM) contains
the Yukawa sector of the theory, $W_Y$, and a Higgs mixing term
with coefficient $\mu$,
\beq
W = W_Y + \mu \H_u \times \H_d,
\label{superpot}
\eeq

\noi
where $\H_u$  and $\H_d$ are Higgs doublet superfields, and where
$\mu$ can be complex.
The low energy supergravity (SUGRA) parametrization of the soft breaking
\Lagr\
can be written in terms of the superpotential and superpartner mass
terms:
\bmulteq
-&{\cal L}_{soft}& = \abs{m_i}^2 \abs{\varphi_i}^2 +
\Bigl( {1 \over2} \sum_\lambda\tilde m_\lambda   \lambda \lambda  + \nonumber\\
&&     A m_0^* \left[ W_Y\right]_\varphi  +
       B m_0^* \left[ \mu \H_u \times \H_d\right]_\varphi  + h.c.\Bigr),
\label{L soft}
\emulteq

\noi
where $\varphi_i$ are the scalar superpartners, $\lambda$ are the
gauginos, and $[\ ]_\varphi$ means take the scalar part.
Like $\mu$, the soft breaking parameters
$A$, $B$, and $m_0$,  and the
gaugino masses $\tilde m_\lambda$, can all be complex.
These parameters contribute to \dn\ at the order of
$10^{-22} \tilde\varphi / \tilde M^2$\ecm, where $\tilde \varphi$ is a
combination of the phases of the parameters, and $\tilde M^2$ is
a combination of superpartner masses, normalized to the weak scale.
The only known ways to make such a large \dn\ compatible with the experimental
upper bound
are to fine-tune the phase $\tilde \varphi$ to order  $10^{-3}$;
have superpartner masses of order a few TeV;
or somehow require all the phases to naturally be zero
\cite{SUSYdn}. Both the first and second approach eliminate much of the
attractiveness of SUSY.  For example, having large superpartner
masses eliminates the possibility of radiative breaking
of $SU_2\times U_1$, which was one of the major successes of SUSY.
Losing this is especially undesirable now that the top mass is large
enough to make it work.

In this letter we consider what {\it natural} amount of CPV is
expected in SUSY theories.  By this we mean the
size of CPV observables one expects in a SUSY theory
constructed such that there are no fine-tunings of phases,
parameters,  or mass scales in order
to satisfy the empirical upper bound on \dn.
Since all the SUSY phases come into $\tilde \varphi$,
these criteria force the parameters $A$, $B$, $m_0$ and $\tilde m_\lambda$
to be real \cite{SUSYdn}.
However, we will show that $\mu$ does {\it not} have
to be real in a large class of models.

We do not propose any explanation for
why the other parameters are real,
but merely accept that to have a SUSY theory without fine-tunings,
these conditions must
somehow be satisfied,   
given the phenomenological and theoretical constraints.
If fine-tunings such as large sparticle masses or phase cancellations turn
out to be important, our arguments may or may not be relevant.
For the remainder of this letter, we simply
take $A,\ B,\ m_0,$ and $\tilde m_\lambda$ to be real;
see \cite{SUSYdn} for a complete treatment of these criteria and
a summary of previous discussions.

Imposing the above criteria means that the MSSM
has no non-SM CPV, thus \dn\ and $d_e$ are tiny, as predicted by the
CKM mechanism.
There are no CKM effects in the renormalization group equations of the
SUSY parameters,
and finite CKM effects are tiny \cite{SUSYdn}.
In addition, there would not be enough CPV to explain the observed
baryon asymmetry. With this in mind, we point out a mechanism
by which a moderate amount of CPV can naturally arise in a
{\it non-minimal} SUSY theory
through loop corrections to the Higgs potential.
The idea is that
a phase which is unobservable at tree level can
introduce an observable effective phase
through loop effects.  But the effective phase
will always be smaller than a tree level
observable phase because of the usual factors of suppression
associated with loops.
Such a phase can make moderate
contributions to \dn, $d_e$, and may be useful in explaining the
baryon asymmetry.
However it requires non-minimal extensions to the MSSM,
and so its consideration will have important effects upon SUSY model building.

\section{Source of \dn}

All \susyic\ contributions to $d_n$ come from the mass matrices
of squarks and gauginos---if the mass
matrices can all be made real, the SUSY contribution to $d_n$ disappears.
If they are complex, the gaugino-squark-quark couplings become
complex and contribute to \dn\ through loop diagrams \cite{hist dn}.
Let us write the down squark mass matrix
in a partially diagonalized basis
\cite{foot rad corr}:
\beq
\pmatrix{ {\mu_{dL}}^2 {\bf 1} + \hat M_D^2 &
(A^* m_0 - \mu \, v_u/v_d) \hat M_D\cr
(A^* m_0 - \mu \, v_u/v_d)^* \hat M_D&
{\mu_{dR}}^2 {\bf 1} + \hat M_D^2\cr } , \label{MDsqk}
\eeq

\noi
where $\hat M_D$ is the diagonal, real,
$N_F \times N_F$ quark mass matrix (where $N_F$ is the number of families),
and ${\mu_{q\, L,R}}^2 \sim \abs{m_{3/2}}^2$.
As mentioned above, we take $A,\ B,\ m_0$ and $\tilde m_\lambda$ to be
real to avoid fine-tuning.
The only remaining possible phases in the squark mass matrices
are those of $\mu$, and the
vacuum expectation values (VEVs; see \pref{defofphis}).

We write the chargino mass matrix, $M_{\chi^+}$,
\beq
\pmatrix { \tilde m_{W} & g_2v_u^* \cr
            g_2v_d      & \mu  \cr} ,
\label{Mchargino}
\eeq

 \noi
in the basis of \cite{HaK},
where $\tilde m_{W}$ is
the $SU_2$ soft breaking gaugino mass, and
$g_2$ is the $SU_2$ coupling constant.

The neutralino mass matrix $M_{\chi^0}$ is (see \cite{HaK}):
\beq
\pmatrix{
 \tilde m_{B} &                   0 & -g_1v_d/\sqrt2 & g_1v_u^*/\sqrt2\cr
                   0 & \tilde m_{W} &  g_2v_d/\sqrt2 &-g_2v_u^*/\sqrt2\cr
      -g_1v_d/\sqrt2 &       g_2v_d/\sqrt2 &              0 &           -\mu
\cr
     g_1v_u^*/\sqrt2 &    -g_2v_u^*/\sqrt2 &           -\mu &              0
\cr
                                           },
\label{Mneutralino}
\eeq

\noi
where $g_1$ is the $U_1$ coupling constant, and
$\tilde m_{B}$ is the $U_1$ gaugino mass.
Most references do not keep track of the phases of the VEVs in
\pref{Mchargino}--\pref{Mneutralino}.
This is undoubtedly due to the fact that the tree level MSSM does not allow for
spontaneous \CPvio\ \cite{GaH,Maekawa}, so most authors have assumed
$v_u$ and $v_d$ are real.
As we will see, there can be a  relative phase between the VEVs
in some cases, so that it is important to use our form
for $M_{\chi^+}$ and $M_{\chi^0}$.


\section{Tree Level Cancellation}

To see why the phase of $\mu$ does not actually contribute to \dn,
we present two arguments. The first relies upon redefinition of the phases of
Higgs superfields.
If we redefine the phases of the Higgs and higgsino
($H_u$ and $\psi_{H_u}$),
it turns out that the $F$ terms get rotated by the same amount,
so that the procedure is equivalent to redefining the phase of the superfield
$\H_u$.  If the soft terms come from the superpotential, as in \pref{L soft},
the phase of $\mu$ gets rotated away too.
In the MSSM, the only new phase introduced by this rotation
is in the Yukawa couplings in $W_Y$.
Since those can absorb an arbitrary phase,
we can define away the phase of $\mu$ without loss of generality,
leaving no CPV in the MSSM \cite{foot provided}, other than the CKM phase.
This is true in the low energy SUGRA parametrization of the MSSM,
and in many extensions of the MSSM.

We also give an alternate
derivation of this result from minimizing the Higgs potential.  In some ways
it is more transparent, and we will need the results in the next section.
We use two doublets of the same hypercharge,
\beq
\phi_1 = {H_d}^c \rightarrow \plet{0 \cr v_d \cr } ,\
\phi_2 = H_u \rightarrow \plet{0 \cr v_u \cr } ,
\label{defofphis}
\eeq

\noi
which defines the two VEVs $v_u$ and $v_d$.
The most general renormalizable $SU_2 \times U_1$ invariant scalar potential
for two Higgs doublets $\phi_1$ and $\phi_2$ is
\cite{GaH,Maekawa,Pomarol one}
\bmulteq
& 
V=m_1^2 \abs{\phi_1}^2 + m_2^2 \abs{\phi_2}^2 - (\mu_{12}^2 \phi_1^\dagger
\phi_2 + h.c.) +
\lambda_1 (\phi_1^\dagger \phi_1)^2 \nonumber\\
& 
 + \lambda_2 (\phi_2^\dagger \phi_2)^2 +
\lambda_3 (\phi_1^\dagger \phi_1)(\phi_2^\dagger \phi_2) +
\lambda_4 (\phi_1^\dagger \phi_2)(\phi_2^\dagger \phi_1) +
\nonumber\\
&  
\left[\lambda_5 (\phi_1^\dagger \phi_2)^2 +
      \lambda_6 \abs{\phi_1}^2(\phi_1^\dagger \phi_2) +
      \lambda_7 \abs{\phi_2}^2(\phi_1^\dagger \phi_2) + h.c. \right] .
\label{fullHiggsPot}
\emulteq

\noi
We see that $m_{1-2}$ and $\lambda_{1-4}$ are real.
Further, SUSY predicts that $\lambda_{5-7}$ are zero at tree level
\cite{GaH,Maekawa},
so that only $\mu_{12}^2$   can be complex.
Using \pref{L soft}, \pref{defofphis}, and \pref{fullHiggsPot} we have,
%
$\mu_{12}^2 = Bm_0^*\mu$ ,
%
and we can define its phase in terms of that of $\mu$ \cite{foot provided}:
%
$\theta_\mu \equiv \Arg\mu =  \Arg\mu^2_{12}$.

Minimizing the Higgs potential in \pref{fullHiggsPot}, one
condition we obtain is
%
$\xi' \equiv \xi + \theta_\mu = 0$,
%
which means that the relative phase between the VEVs,
$\xi \equiv \Arg{v_u \over v_d}$, is non-zero.  This phase
can be rotated away everywhere
in the gauge and matter sectors {\it except}
in the squark and gaugino mass matrices.
We consider the effects of $\xi'=0$ on each.

The essential point is that the
minimization condition $\xi' = 0$ implies that $\mu\, v_u/v_d$
is real, which means that
the squark mass matrix in \pref{MDsqk}
is also real \cite{foot provided}.
Thus the phases of $\mu$ and the VEVs cancel, so
there is no squark mass matrix contribution to \dn,
even if $\mu$ is complex.

A necessary condition for this cancellation is that
the Higgs scalar mixing term in ${\cal L}_{soft}$ has the same
phase as the Higgs superfield mixing term in the superpotential;
we must have $\mu_{12}^2 \sim \mu$ {\it and} the no fine-tuning criteria which
implies $B m_0^*$ is real.  If there are contributions
to $\mu_{12}^2$  which are not proportional to the phase of $\mu$,
the cancellation goes away.
Note that if $\mu$ is zero, the phase of $\mu_{12}^2$, whatever its source,
can be rotated away.

The cancellation in the gaugino mass matrices is more subtle.
Contributions to \dn\ can come from phases in the
gaugino-squark-quark vertices, which are introduced by the unitary matrices
which diagonalize $M_{\chi^+}$ and $M_{\chi^0}$.  But there is also
a phase in the higgsino-squark-quark part of the couplings
from the VEVs.  In gaugino-squark-quark vertices,
$\psi_{H_d}^-$ ($\psi_{H_u}^+$) higgsinos are always accompanied by
$1/v_d^*$ ($1/v_u$).
These phases can be rotated into the definition
of $\psi^-_{H_d}$ and $\psi^+_{H_u}$,
so that the weak basis gaugino-squark-quark couplings are real.
This redefinition of the weak states also changes the gaugino mass
matrices.  It turns out that the  only phase in the
rotated $M_{\chi^+}$ and $M_{\chi^0}$  is $\xi'$, and thus the minimization
condition implies that $M_{\chi^+}$  and $M_{\chi^0}$ are real!
Remarkably, by having the $\mu_{12}^2 \sim \mu$,
the VEVs are aligned so the mass matrices
\pref{MDsqk}--\pref{Mneutralino} are all real, and there is no
SUSY contribution to \dn\ at tree level even if $\mu$ is complex.

\section{A Loop Induced Observable Phase}

One might be worried that after inclusion of one loop effects
that one could no longer rotate away
the phase of $\mu$, since there are
new terms which involve $H_u$.
This cannot happen because at tree level all
vertices are independent of the phase of $\mu$, so there is no way for
it to reappear at one loop.  However, it is possible for a phase
which makes negligible contribution to CPV in the tree level
Lagrangian to appear more prominently at one loop.

After one loop corrections to $V$, there will be non-zero contributions
to $\lambda_{5-7}$ and $\mu_{12}^2$.  In the natural MSSM (satisfying
our criteria), these
contributions (call them $\delta\lambda$) will be real.
Suppose that our theory contains some new
 complex parameter which does not lead to
large CPV through tree level vertices.  Suppose further
that the complex parameter appears in
vertices which involve Higgs fields.
Then the parameter could introduce complex $\delta \lambda$'s through loops.
The VEVs would get a relative phase,
and that could in turn introduce observable CPV effects.

To see how this works, let us consider the one loop Higgs potential
with arbitrary coefficients $\delta\lambda$ added to the tree level $V$.
Let us also use our results from the last section to set
$\xi=\xi'=\theta_\mu =0$ at tree level.
At one loop, the minimization point will shift to
\bmulteq
&&\sin\xi \simeq {v^2\over \abs{\mu_{12}^2}}\,
\Im\Bigl[ \delta\lambda_5 \sin 2\beta \pm \nonumber \\
&&( - \delta\mu_{12}^2/v^2
+ \delta\lambda_6 \cos^2\beta + \delta\lambda_7 \sin^2\beta) \Bigr]
\label{sinxi}
\emulteq

\noi
where the $+$ ($-$) corresponds to $\xi$ near $0$ ($\pi$), and
$\tan\beta \equiv \abs{v_u/v_d}$.  We have assumed
$\mu_{12}^2 >> \delta\lambda v^2$ for simplicity.

This new induced phase $\sin\xi$ contributes to both \dn\ and
$d_e$ through the sfermion and gaugino mass matrices.
Let us define the phase of the quantity in brackets in \pref{sinxi}  to
be $\theta_{\delta\lambda}$, and take the magnitude of these loop corrections
to be of order $10^{-3}$.
If we take
$B \sim .5$, colored superpartners $\sim 300$GeV, sleptons $\sim 175$GeV, and
all other superpartners $\sim 100$GeV,
we estimate that
\bmulteq
&&d_n \sim 10^{-26} \tan\beta \, \sin\theta_{\delta\lambda} \, {\rm e\, cm} ,
\label{dn one loop}\\
&&d_e \sim .6\times 10^{-27} \tan\beta \, \sin\theta_{\delta\lambda} \,
{\rm e\, cm} .
\label{de one loop}
\emulteq

\noi
These estimates do depend upon the parameters and the mass scales in the
theory,
%
but the point is that the contributions entering at one loop
are naturally much smaller than
those from SUSY phases which contribute through tree level vertices.
Note that this effect will not work in the MSSM \cite{foot provided}
because the CKM contribution to
$\theta_{\delta\lambda}$ is negligible.
We must look beyond the MSSM.

As an  example, let us add a gauge singlet superfield $\N$ to the minimal
model \cite{foot MNMSSM}.
Suppose  we replace the superpotential in \pref{superpot} with
\beq
W = W_Y + \mu \H_u \times \H_d +
h \N \H_u \times \H_d + {1\over 3} k \N^3,
\eeq

\noi
where $h$ and $k$ are complex, and $N$ does {\it not} get a VEV.
By redefinition of the phase
of $\N$ we can make $h$ real, leaving $k$ arbitrarily complex.
With $\vev{N}=0$ the phase of $k$, $\theta_k$, does not appear in any vertices
outside the Higgs sector at tree level.
Some scalar-pseudoscalar mixing is induced,
but one loop contributions to \dn\ and $d_e$ from this are negligible
\cite{NGaGENG}, though two loop diagrams may be larger
\cite{BZ GV}.
However, one loop corrections to $V$ can depend upon $\theta_k$, and thus
perhaps $\sin\theta_{\delta\lambda} \sim \sin\theta_k$, which
could be of order unity.  Thus the phase $\theta_k$, which at present
is probably unobservable at tree level, can give \dn\ and $d_e$ near their
current limits through loop effects.

\section{Remarks}

We have shown that it is not necessary for $\mu$ to be real
for a SUSY theory to satisfy the bound on \dn\ without fine-tunings.
The soft breaking parameters $A$, $B$, $m_0$ and $\tilde m_\lambda$ must
be real or a new mechanism must be found to give them phases at the
$10^{-3}$ level.
A MSSM which satisfies
the no fine-tuning criteria gives no
new CPV beyond the SM CKM phase, and thus \dn\ and $d_e$ are negligible.
There may still be observable CPV effects due to SUSY CKM contributions
to various processes, notably in B decays \cite{Bigi B}.
We showed that non-minimal SUSY models
can potentially give a loop induced observable phase,
which can give \dn\ and $d_e$ near their current bounds.

Note that the cancellation of the phase of $\mu$ depends
crucially on the assumption that $\mu_{12}^2$ is
proportional to the phase of $\mu$.
This will not be true if $\mu_{12}^2$ has sources other than the
superpotential, $e.g.$ $\mu_{12}^2$ is put in by hand.
Knowing that the phase of $\mu$ can be rotated away
is important because
SUGRA models do not usually generate superpotential parameters, so
explaining why the phase of $\mu$ was zero might have proven much
harder than for the other parameters.

In the MSSM a non-zero $d_n$ or $d_e$ near the present limits  would
probably have pointed away from SUSY theories, since there would have to
have been some fine-tuning in the theory.  For SUSY believers a non-zero
observation of $d_n$ or $d_e$ would point toward non-minimal SUSY models
which allow CPV to enter at one loop as we have described
above, or require a new mechanism to give the soft breaking parameters a
tiny phase.

There are also SUSY contributions to \dn\ from
a three gluon operator \cite{three gluon},
which are probably smaller than the quark EDM contribution \cite{arnowitt}.
If the no fine-tuning criteria we have used is satisfied,
this operator will give only a small numerical correction to \dn,
proportional to the phase $\theta_{\delta\lambda}$.
Thus \pref{dn one loop}--\pref{de one loop} still
reflect the natural level of CPV
possible in a non-minimal SUSY model.

It may be possible to have {\it spontaneous} \CPvio\ in \susyic\ theories at
one
loop \cite{Maekawa}, though that may conflict with Higgs mass limits
\cite{Pomarol one}.
That could be circumvented by extending the MSSM with a singlet $\N$ and
allowing it a complex VEV, giving spontaneous \CPvio\ at tree level
\cite{Pomarol two}.
In either case a fine-tuning is probably required of the VEV phase $\xi$
to keep \dn\ below
the experimental bound; in fact we argue that spontaneous \CPvio\ generally
requires {\it more} fine tuning than hard \CPvio.
One of the minimization conditions necessary
for non-zero spontaneous CPV is
\beq
\cos\hat\xi = {X \over Y} \simeq 1 - {1\over 2} \hat\xi^2,
\eeq

\noi
where $\hat\xi$ is $|\xi|$ or $|\xi - \pi |$, and $X$ and $Y$ are
some combination of loop integrals, parameters, and
VEVs. We need $\hat\xi$ to be small to satisfy the bound on \dn,
which we can achieve only if $\Delta \equiv (Y-X)/Y$ is of order $\hat\xi^2$.
For example, if we need $\hat\xi\sim 10^{-3}$, then $\Delta$ must
be fine-tuned to be of order $10^{-6}$, which is completely unacceptable.
Thus SUSY theories of spontaneous CPV may have
even more trouble in justifying the small experimental bound on \dn\
than hard CPV.
However, it should be fairly easy in most models to require there be
{\it no} spontaneous CPV, without any fine-tunings.

Finally we note that the loop mechanism of section IV
could be very important to the
CPV aspects of the problem of the baryon asymmetry.
A recent interesting model of electroweak baryogenesis used
CPV from the Higgs scalar mixing coefficient $\mu_{12}^2$
\cite{Turok Zadrozny}.
It was pointed out \cite{Dine BAU} that  $\mu_{12}^2$
can be rotated out of the Higgs
potential, but the resulting phase in the gaugino mass matrices
was then used by \cite{Cohen Nelson}.  They found that with the small phase
allowed by the limit imposed by \dn, there is probably
sufficient CPV for the baryon asymmetry.
Our results change these conclusions in two ways:  At tree level, there
is {\it no} phase in the gaugino mass matrices \cite{foot provided},
and no way for $\theta_\mu$ to cause CPV.
At one loop in non-minimal SUSY models, there can be an effective phase $\xi$
introduced into the
gaugino mass matrix of order
$\xi \sim 10^{-3} \theta_{\delta\lambda}$.
Although this is a large suppression, $\theta_{\delta\lambda}$ can be of order
unity and so $\xi$ should generate the same level of CPV
as the phase used in  \cite{Cohen Nelson}, which was bounded by \dn\ anyway.

{}From the standpoint of explaining the baryon asymmetry or a non-zero \dn\
or $d_e$, a loop induced observable phase
provides an attractive alternative to
the fine-tuning needed in the MSSM.
If \dn\ or $d_e$ is observed in the near future, or the baryon asymmetry
is generated at the weak scale, it will either force believers in
SUSY toward non-minimal models, or require a way of somehow naturally
generating
phases of order $10^{-3}$.


\centerline{ACKNOWLEDGEMENTS}

We appreciate helpful comments from J.-M. Fr\` ere, L. Roszkowski,
A. Nelson, and particularly from H. Haber.




\begin{references}
\bibitem{foot strong}We do not consider strong CPV effects in this letter.
\bibitem{foot rad corr}This matrix is written at tree level.  We show in
\cite{SUSYdn} that radiative corrections do not change
our conclusions. The up squark effects parallel those of the down squarks.
\bibitem{foot provided}Provided that the SUSY theory satisfies the
criteria that $A,\ B,\ m_0$ and $\tilde m_\lambda$ are real.
\end{references}
\end{document}